\documentclass{ws-procs9x6}

\newcommand{\newangle}{\sphericalangle}

\begin{document}
\title{Two-hadron single target-spin asymmetries: \\ first measurements by HERMES}
\author{\underline{P.B. van der Nat}, K. Griffioen \\ (on behalf of the HERMES collaboration)}
\address{Nationaal Instituut voor Kernfysica en Hoge-Energiefysica (NIKHEF),
\\ P.O. Box 41882, 1009 DB Amsterdam, The Netherlands}
\maketitle
\abstracts{
Single target-spin asymmetries in semi-inclusive two-pion production were measured for the first time by the
HERMES experiment, using a longitudinally polarized deuterium target. These asymmetries relate to
the unknown transversity distribution function $h_1(x)$ through, also unknown, interference fragmentation functions. The
presented results are compared with a model for the dependence of one of these interference
fragmentation functions on the invariant mass of the pion pair.
}
\section{Introduction}
Of the three leading-twist quark distributions, the quark number density,
the quark helicity and the quark transversity distribution, only the latter
one is so far unmeasured. The main reason for this is its chiral-odd
nature which requires a second chiral-odd object to couple to the transversity
distribution in order to make it accessible to measurements. One candidate for such a chiral-odd
fragmentation function is the so-called Collins fragmentation
function appearing in pion lepton production. This has been studied by the HERMES experiment
using a longitudinally polarized target\cite{HERMES_longpol}, and recently using a transversely
polarized target\cite{HERMES_transpol}.

Another way of accessing transversity is offered by interference fragmentation
functions, which appear in single target-spin asymmetries in two-pion semi-inclusive deep-inelastic
scattering (DIS). One of the advantages of this method is that the azimuthal moment of the asymmetry is directly proportional
to products of distribution and fragmentation functions, whereas in the case of one-hadron
semi-inclusive DIS, these products are convoluted with the transverse momentum of the detected
hadron. Although the interference fragmentation functions themselves are as yet unknown, they can be cleanly measured in 
$e^+e^-$ experiments, such as Belle\cite{Belle_Ralf} and Babar.

\section{Single Spin Asymmetry}
\begin{figure}[t]
\includegraphics[width=6.5cm]{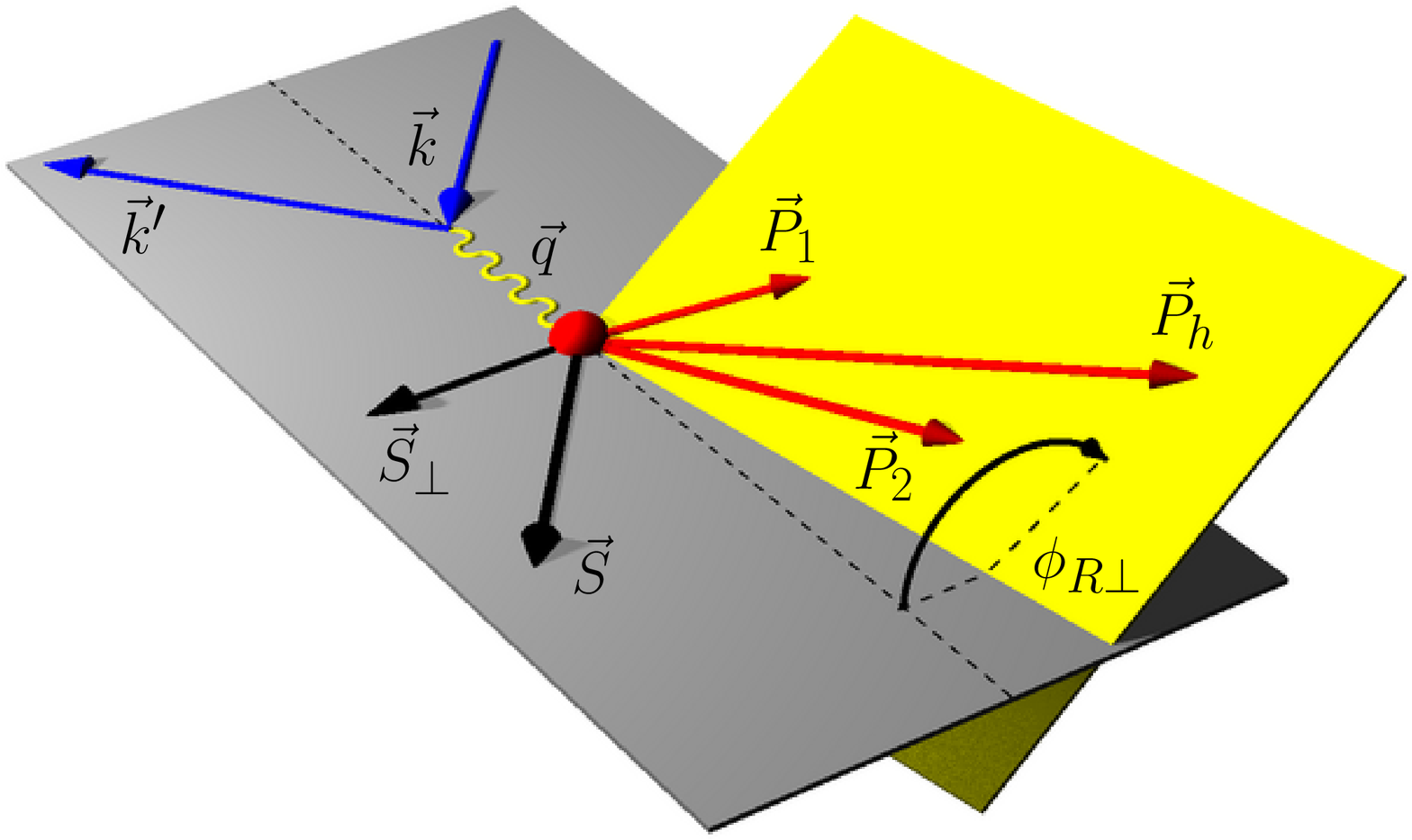}
\rule{.5cm}{0pt}
\includegraphics[width=4cm]{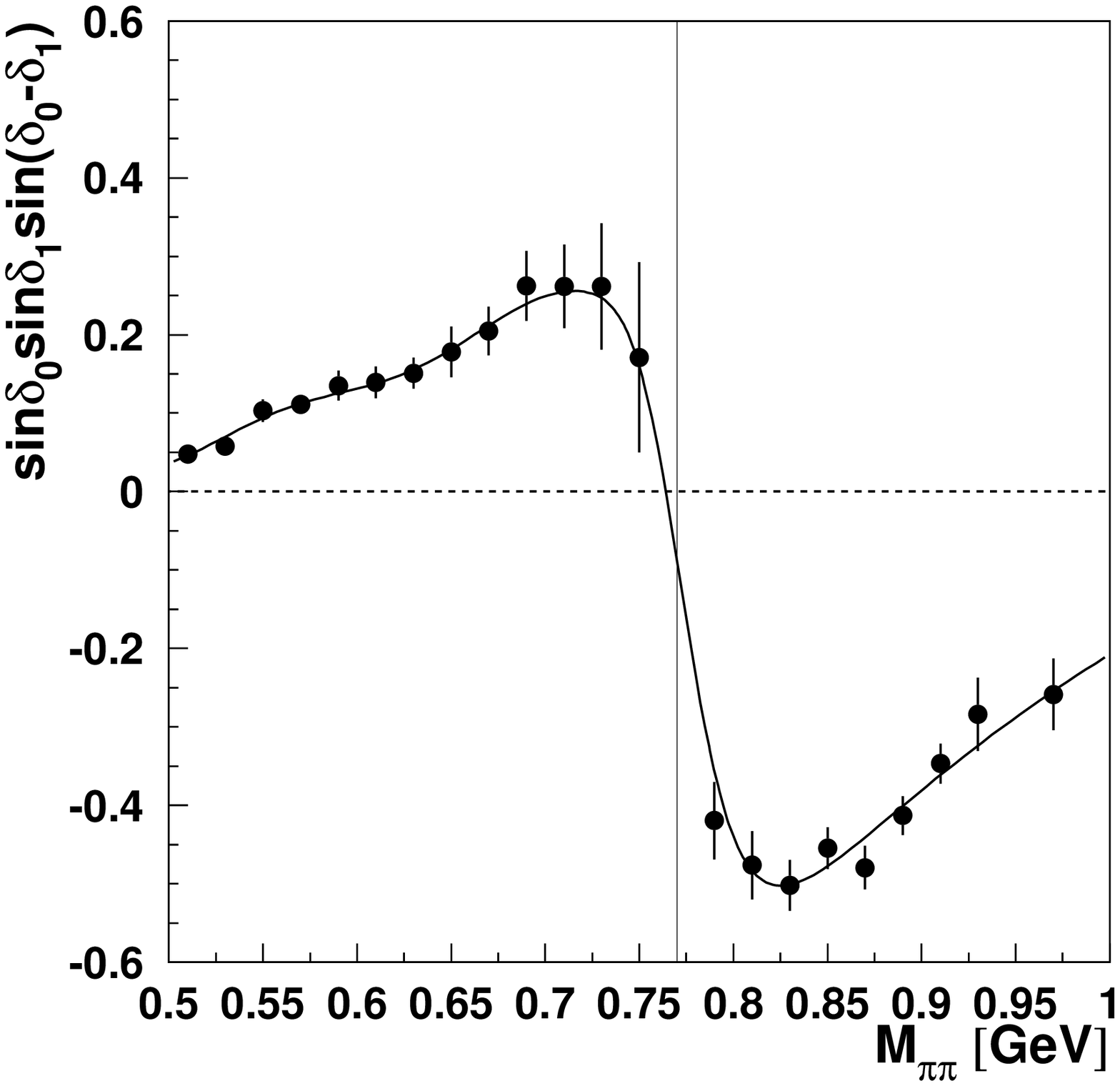}
\caption{Left: kinematic planes, where $\phi_{R \perp}$ is the angle between the
plane spanned by the incident ($\vec{k}$) and scattered lepton ($\vec{k}'$) and the plane spanned by the two
detected pions ($\vec{P}_1$ and $\vec{P}_2$, with $\vec{P}_h \equiv \vec{P}_1 + \vec{P}_2$). Right:
the factor  $\sin\delta_0\sin\delta_1\sin(\delta_0-\delta_1)$, where the s- and
p-wave phase shifts ($\delta_0$ and $\delta_1$) were obtained from pion-nucleon scattering
experiments\protect\cite{phaseshifts}. This factor shows the invariant mass dependent part of $H_1^{\newangle,sp}$, as predicted by Jaffe \emph{et al.}\protect\cite{Jaffe}.}\label{fig:azimuthal_angle}
\end{figure}

The transversity distribution can be accessed experimentally by measuring the single target-spin asymmetry,
defined as:
\begin{equation}
A_{UL}(\phi_{R \perp})  =
\frac{1}{|P_L|}\frac{N^{\rightarrow}(\phi_{R \perp})/N_d^{\rightarrow} - N^{\leftarrow}(\phi_{R \perp})/N_d^{\leftarrow}}{N^{\rightarrow}(\phi_{R \perp})/N_d^{\rightarrow} + N^{\leftarrow}(\phi_{R \perp})/N_d^{\leftarrow}   } = \frac{\sigma_{UL}}{\sigma_{UU}}\mathrm{,}
\end{equation}
where $N^{\rightarrow}$ ($N^{\leftarrow}$) is the number of $\pi^+ \pi^-$ pairs detected with
target spin antiparallel (parallel) to the direction of the beam momentum. These numbers are
normalized to the corresponding number of DIS events, $N_d^{\rightarrow}$ and $N_d^{\leftarrow}$, 
respectively and the entire ratio is divided by $P_L$, the longitudinal target polarization. The asymmetry is evaluated as a function of the azimuthal angle $\phi_{R \perp}$, which is shown in Fig. \ref{fig:azimuthal_angle}. 
In the last term, $\sigma_{UL}$ and $\sigma_{UU}$  are the  polarized and unpolarized cross
sections, respectively.
According to Bacchetta \emph{et al.}\cite{Alessandro_DIS2004} $\sigma_{UL}$ can be written at
sub-leading twist as\footnote{Eq. \ref{eq:sigma_UL} was derived using the
Wandzura-Wilczek approximation and is valid at low invariant mass $M_{\pi\pi}$ of the pion pair}:
\begin{equation}
\sigma_{UL} \sim \sum_q e_q^2 \sin \phi_{R \perp} \sin \theta \left[K_1 |\boldsymbol{S}_{\parallel}|  h_L
 - K_2 |\boldsymbol{S}_{\perp}| h_1  \right] \left(  H_1^{\newangle,sp} + H_1^{\newangle,pp} \cos
\theta \right)\mathrm{,}\label{eq:sigma_UL} 
\end{equation}
where $K_1$ and $K_2$ are kinematic factors\footnote{See the article by Bacchetta \emph{et al.}\cite{Alessandro_DIS2004} for the full
expression.} and $\theta$ is the angle between the direction of emission of the
pion pair in its center-of-mass frame and $\vec{P}_h$ in the hadronic frame (see
Fig. \ref{fig:azimuthal_angle}).  
Eq. \ref{eq:sigma_UL} introduces the two-hadron interference fragmentation
functions $H_1^{\newangle,sp}$ and $H_1^{\newangle,pp}$. They decribe the
interference between different production channels of the pion pair: $H_1^{\newangle,sp}$ relates to
the interference between s- and p-wave states and $H_1^{\newangle,pp}$ to the interference between
two p-wave states.
Both functions can be used \emph{separately} to extract information on the transversity
distribution $h_1(x)$. 
In the present analysis the $\sin\phi_{R \perp}$-moment of the asymmetry, $A_{UL}^{\sin \phi_{R \perp}}$ has been
studied, which is only sensitive to $H_1^{\newangle,sp}$, because in evaluating $A_{UL}^{\sin
\phi_{R \perp}}$ the integral is taken over the polar angle $\theta$.

A prediction was given by Jaffe \emph{et al.}\cite{Jaffe} for the invariant-mass behavior of $H_1^{\newangle,sp}$ in
terms of s- and p-wave phase shifts. Fig. \ref{fig:azimuthal_angle} shows that
according to this model the asymmetry would change sign approximately at the $\rho^0$ mass. Note,
however, that this model does not predict the size or sign of the asymmetry.

Two distribution functions appear in Eq. \ref{eq:sigma_UL}: the transversity distribution
$h_1$ and the subleading-twist function $h_L$, which is related to $h_1$ through a
Wandzura-Wilzcek relation. The contribution of
these functions is proportional to the target polarization components transverse
($\boldsymbol{S}_{\perp}$) and parallel ($\boldsymbol{S}_{\parallel}$) to the virtual
photon direction, respectively. In the data presented here the value of $\boldsymbol{S}_{\perp}$
increases from  3\% of $\boldsymbol{S}_{\parallel}$ at low $x$ to 9\% at high $x$. 
For the present analysis data were taken during the period 1998-2000 with a longitudinally polarized
deuterium (gas) target. The average target polarization was 0.84 $\pm$ 0.04.
\section{Results}
\begin{figure}[t]
\center
\vspace*{-.5cm}
\includegraphics[width=10cm]{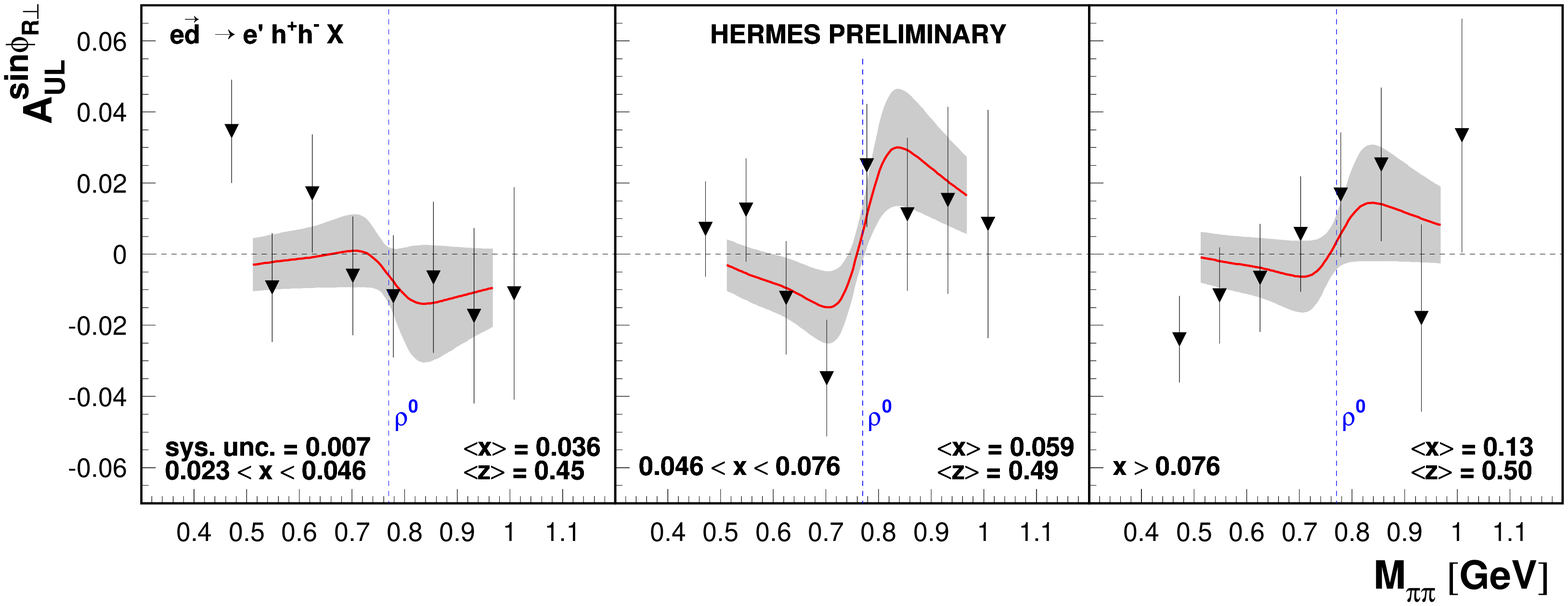}

\vspace*{-.5cm}
\includegraphics[width=7cm]{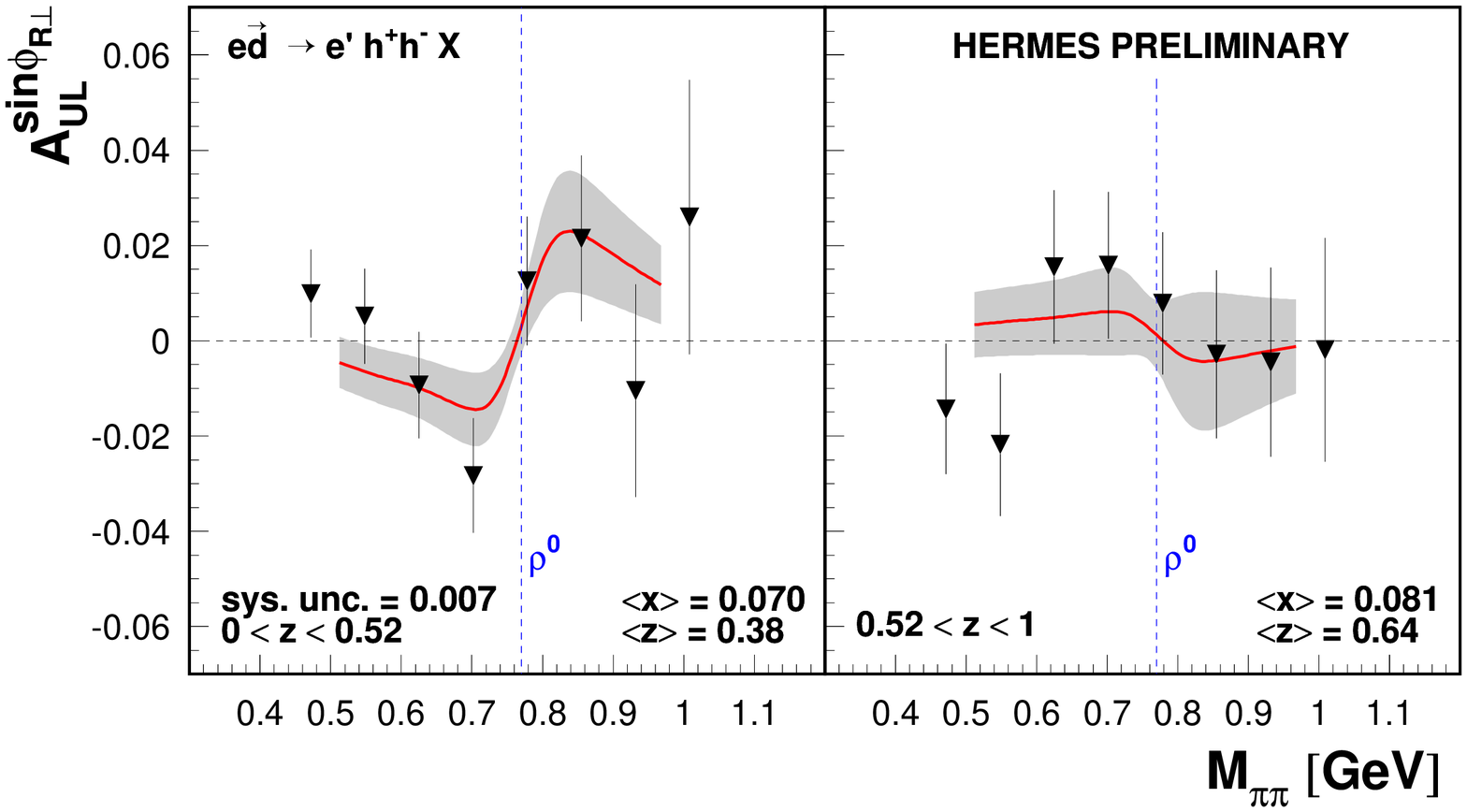}
\vspace*{-.3cm}
\caption{The asymmetry-moment $A_{UL}^{\sin \phi_{R \perp}}$ as a function of the invariant mass
$M_{\pi\pi}$ for different $x$-bins (top) and $z$-bins (bottom).}
\label{fig:results}
\end{figure}                                                                                                    
In Fig. \ref{fig:results} the $\sin \phi_{R \perp}$-moment\protect\footnote{Using the definition of $\phi_{R \perp}$ shown in Fig. \ref{fig:azimuthal_angle}, $A_{UL}^{\sin \phi_{R \perp}}$ will differ by a sign, compared with the situation where the azimuthal angle is defined according to 	the Trento conventions\protect\cite{Trento_conventions}.} $A_{UL}^{\sin \phi_{R \perp}}$ is plotted versus the
invariant mass of the pion pair\footnote{For these preliminary results, all hadron types were
analyzed assuming they were pions. The corresponding uncertainty is not included in the quoted systematic
error.} $M_{\pi\pi}$ in panels of increasing $x$ (= $Q^2/(2M\nu)$) and $z$ ($z \equiv E_{\pi\pi}/\nu$).
The size of the asymmetries is on the order of a few percent. For all panels the asymmetries are
not inconsistent with zero given the size of the statistical errors. No significant $x$- or
$z$-dependence is observed. 
The shape of the asymmetries versus the invariant mass has been compared with the model prediction
shown in Fig. \ref{fig:azimuthal_angle}. This was done by fitting the following function to the data:
\begin{equation}
f(M_{\pi\pi}) = c_1 \mathcal{P}(M_{\pi\pi}) + c_2 \label{eq:fitfunc_jaffe}
\end{equation}                                                                                         
where $\mathcal{P}(M_{\pi\pi})$ contains the invariant-mass dependence from the model prediction and
$c_1$ and $c_2$ are free parameters of the fit. The resulting 
curves are included in Fig. \ref{fig:results}. These curves show that in all panels,
the results are consistent with the model. In the mid-$x$ and low-$z$ region the data
give a hint for a sign change of the asymmetry at the $\rho^0$ mass.

Starting in 2002, HERMES has been taking data with a transversely polarized hydrogen target,
with an average polarization of 0.78 $\pm$ 0.04, which can result in much larger asymmetries.
Data-taking will continue until the summer of 2005. The analysis of these data is ongoing and first results are
expected in the near future.


\begin{thebibliography}{0}
\bibitem{HERMES_longpol}  A. Airapetian \emph{et al.} (HERMES), {\it Phys. Lett.} B {\bf 562}, 182 (2003).
\bibitem{HERMES_transpol} A. Airapetian \emph{et al.} (HERMES), {\it Phys. Rev.  Lett.} (in press), hep-ex/0408013.
\bibitem{Belle_Ralf} R. Seidl (Belle) these proceedings.
\bibitem{Trento_conventions} A. Bacchetta \emph{et al.}, hep-ph/0410050 (2004)
\bibitem{Jaffe} R.L. Jaffe, X. Jin, and J. Tang. {\it Phys. Rev. Lett.} {\bf 80}, 1166 (1998).
\bibitem{Alessandro_DIS2004} A. Bacchetta and M. Radici, {\it Proceedings of DIS'2004}
(2004), hep-ph/0407345.
\bibitem{Alessandro_PRD67} A. Bacchetta and M. Radici, {\it Phys. Rev.} D {\bf 67}, 094002 (2003), hep-ph/0407345.
\bibitem{phaseshifts} P. Estabrooks and A. Martin, {\it Nucl. Phys.} {\bf B79},
301 (1974).
\end{thebibliography}
\end{document}